# Interplay of Structural Chirality, Electron Spin and Topological Orbital in Chiral Molecular Spin Valves

Yuwaraj Adhikari[1,*], Tianhan Liu[1,*], Hailong Wang[2], Zhenqi Hua[1], Haoyang Liu[1], Eric Lochner[1], Pedro Schlottmann[1], Binghai Yan[3,+], Jianhua Zhao[2,+], Peng Xiong[1,+]

[1] Department of Physics, Florida State University, Tallahassee, Florida 32306, USA

[2] State Key Laboratory of Superlattices and Microstructures, Institute of Semiconductors, Chinese Academy of Sciences, Beijing 100083, China

[3] Department of Condensed Matter Physics, Weizmann Institute of Science, Rehovot, Israel

* These authors contributed equally.

[+] emails: binghai.yan@weizmann.ac.il, jhzhao@semi.ac.cn, pxiong@fsu.edu

Abstract: **Chirality has been a property of central importance in chemistry and biology for more than a century, and is now taking on increasing relevance in condensed matter physics. Recently, electrons were found to become spin polarized after transmitting through chiral molecules, crystals, and their hybrids. This phenomenon, called chirality-induced spin selectivity (CISS), presents broad application potentials and far-reaching fundamental implications involving intricate interplays among structural chirality, topological states, and electronic spin and orbitals. However, the microscopic picture of how chiral geometry influences electronic spin remains elusive. In this work, via a direct comparison of magnetoconductance (MC) measurements on magnetic semiconductor-based chiral molecular spin valves with normal metal electrodes of contrasting strengths of spin-orbit coupling (SOC), we unambiguously identified the origin of the SOC, a necessity for the CISS effect, given the negligible SOC in organic molecules. The experiments revealed that a heavy-metal electrode provides SOC to convert the *orbital* polarization induced by the chiral molecular structure to *spin* polarization. Our results evidence the essential role of SOC in the metal electrode for engendering the CISS spin valve effect. A tunneling model with a magnetochiral modulation of the potential barrier is shown to quantitatively account for the unusual transport behavior. This work hence** **produces critical new insights on the microscopic mechanism of CISS, and more broadly, reveals a fundamental relation between structure chirality, electron spin, and orbital.**

Helical textures and monopole-like chirality in electronic structures of topological materials have given rise to a plethora of emergent phenomena characterized by unusual interplays between electronic charge, spin, and orbital [1–4]. More recently, a parallel phenomenon in real space, in which structural chirality induces electron spin polarization in the direction of their momentum, has received increasing attention [5–7]. The effect, termed chirality-induced spin selectivity (CISS), was first evidenced by Mott polarimetry of photoelectrons from a nonmagnetic (NM) Au electrode through a self-assembled monolayer (SAM) of short synthetic molecules of dsDNA [8]. Since then, CISS has been observed in a variety of chiral molecular systems including macro [8–13] and small [14,15] molecules, supramolecular polymers [16], metal-organic frameworks [17], and hybrid organic-inorganic perovskites [18,19] and artificial superlattices [20,21], via a host of electrical, optical, and electrochemical probes [22–25]. More broadly, CISS is shown to effect enantio-selective chemical reactions [26] and facilitate enantiomer separation [27], and the adsorption of chiral molecules on the surface of a conventional superconductor was reported to induce unconventional superconductivity [28,29]. All these experiments suggest a highly consequential interaction between molecular structural chirality and electronic spin, which carries profound and broad implications.

Despite increasing preponderance of experimental results and a great deal of theoretical efforts, the microscopic origin and physical mechanisms behind CISS remain open questions [30,31]. A central unsettled issue is the role of spin-orbit coupling (SOC) in the chiral media. SOC is a necessary element in the emergence of spin polarization in *NM* materials in general. Specifically, it is an essential ingredient in most theoretical models of CISS, whereas the SOC in the molecular materials is too weak to account for the experimentally significant spin selectivity at room temperature [32]. In order to overcome this difficulty, a number of theoretical approaches were proposed, based primarily on spin dependent scattering and tight-binding models [33–41]. The approaches have targeted at amplification of SOC, either its *value*, to account for the experimentally observed CISS spin polarization by introducing various factors such as density of scattering centers [36], dephasing [37,42], and environmental nonunitary effects [41], or its *effect*, through electron-electron correlation [43], exchange interactions [44], vibrational and polaronic effects [45,46], frictional dissipation [47], and Berry force [48].

An alternative approach rids of reliance on SOC in chiral molecules altogether [39,49,50]. Gersten *et al*. [39] introduced the concept of "induced spin filtering": A selectivity in the transmission of the electron orbital angular momentum can induce spin selectivity in the transmission process, provided that there is strong SOC in the substrate supporting the chiral SAM. This proposal, however, was questioned because CISS was observed in photoemission experiments in which the substrates have negligible SOC [11,51]. Liu *et al.* [49] noted an important difference between the manifestations of the CISS effect in photoemission setups [8,15] and transport in molecular junctions [52,53]: The former measures the "global orbital angular momentum" that includes both the orbital and spin angular momenta, whereas the latter probes spin polarization exclusively. Physically, the model suggests that chiral molecules act as an *orbital* filter rather than a *spin* filter, and the SOC in the metal electrode converts the orbital polarization into spin polarization, thus producing CISS without the need for any SOC in the molecules (see the illustration in Fig. 1a). The orbital polarization effect, which is caused by the orbital-momentum locking —an intrinsic topological property of electronic states in a chiral material [49,54], has much broader relevance beyond CISS; in particular, it presents a new pathway for spin manipulation through atomic structure engineering. So far, however, definitive experimental evidence of the effect is lacking.

One of the most widely used device platforms to detect and utilize spin polarization are spin valves. In a CISS spin valve, the spin polarization of a charge current from the NM electrode through the chiral SAM is analyzed by the magnetic electrode, and the junction conductance is expected to depend on the magnetization direction of the magnetic electrode, resulting in a MC. For CISS studies, a scanning probe rendition of the spin valve, magnetic conductive atomic force microscopy (mc-AFM), has been frequently used [52]. Although its implementation is relatively straightforward, mc-AFM relies on large number of averaging to mitigate the fluctuation and instability. In contrast, thin film-based molecular junctions, in which a chiral SAM is sandwiched between two conducting electrodes, are more conducive to stable and reproducible current-voltage (I-V) and MC measurements. Such devices, however, present significant technical challenges of their own: Pinholes are almost always present in SAMs at device scales (~ µm), and in the cases of two metal electrodes, any direct contact will short out the device.

We recently demonstrated that these complications can be effectively mitigated by using a ferromagnetic semiconductor, (Ga,Mn)As, as the magnetic spin analyzer [53]. The use of the magnetic semiconductor was found to alleviate the shorting problem due to the presence of a Schottky barrier at direct contact with the Au electrode. Moreover, the (Ga,Mn)As was grown by molecular beam epitaxy (MBE) on an (In,Ga)As buffer layer, and the resulting strain from the lattice mismatch produces perpendicular magnetic anisotropy (PMA) [55], namely an out-of-plane magnetization that is collinear with the spin polarization from CISS. These two unique device characteristics enabled observation of spin-valve MC distinctly associated with CISS, and the inherent stability of the platform facilitated a first rigorous determination of the bias current dependence of the MC from CISS [53].

Leveraging this proven device platform, we fabricated and characterized a deliberately chosen set of (Ga,Mn)As/SAM/NM hetero-junctions. The experiments yielded *quantitative* differentiation of the magnitude and bias-dependence of the spin valve conductance in junctions with NM electrodes of contrastingly different SOC strengths (Au versus Al) and SAMs of chiral and achiral molecules. The results revealed a definitive correlation between the magnitude of the CISS spin valve conductance and the SOC strength in the NM electrode: The molecular junctions with Au electrodes exhibit significant MC whose magnitudes depend distinctly on the chirality or length of the molecules; in contrast, in otherwise identical devices with Al electrodes, regardless of the molecule involved the MC are essentially indistinguishable from those of the control samples without any molecules. A model based on magnetochiral modulation of the tunneling barrier potential [56] from orbital polarization is shown to provide quantitative account for both the magnitude and bias dependence of the MC of the two types of junctions. The work unambiguously evidenced the essential role of the contact SOC in generating observable CISS effect in chiral molecular spin valves.

**Chiral molecular spin valve and orbital to spin conversion**

We detect the MC in chiral molecular spin valve devices with a (Ga,Mn)As magnetic electrode and Au or Al normal metal electrode. Figure 1a shows a schematic diagram depicting the molecular junction structure and the physical mechanism for the chirality-induced orbital polarization and subsequent orbital to spin polarization conversion due to the SOC in the NM electrode. Figure 1b are schematics of the device heterostructure and setups for the quasi-four-terminal I-V and conductance measurements. Figure 1c is an SEM image of a junction; the junctions were squares of sizes ranging from $5 \times 5$ µm$^2$ to $15 \times 15$ µm$^2$.

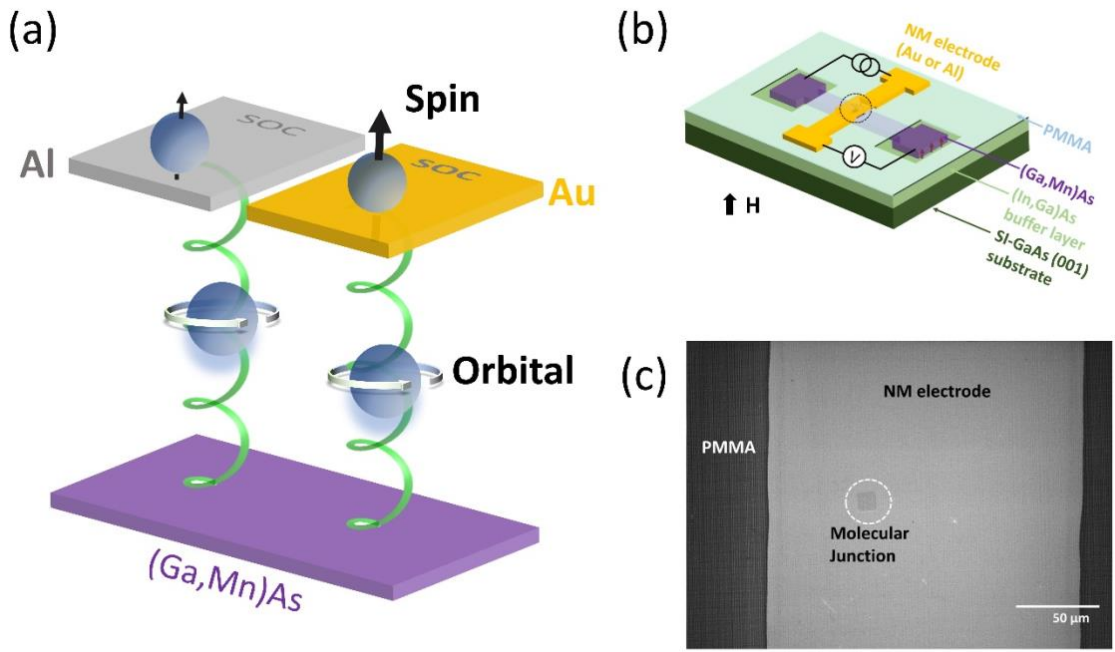

**Fig. 1: The chiral molecular spin valve and orbital to spin conversion mechanism.** (a) Schematic depiction of the mechanism for orbital to spin polarization conversion in a (Ga,Mn)As/chiral molecule/NM spin valve. Chiral molecules induce orbital polarization in passing electrons and subsequently, the electrode SOC converts the orbital to spin (represented by the black vertical arrows on the top electrodes) [49]. (b) Schematic diagram of the device structure along with the junction measurement setup. (c) A top-view scanning electron microscopy image of the junction region (black circle in (b)) in a molecular spin-valve device.

Molecular assembly, the formation of the SAM on (Ga,Mn)As, is a critical step in the device fabrication process. For this work, alpha-helix L-polyalanine (AHPA-L) and L-cysteine served to compare chiral molecules of different molecular lengths, whereas 16-mercaptohexadecanoic acid (MHA) and 1-octadecanethiol (ODT) were used as achiral molecules of similar length but with polar and nonpolar terminal groups respectively [57]. The total molecular length of AHPA-L is 5.4 nm, whereas L-cysteine is around 4 Å, and those of MHA and ODT are 2.4 nm and 2.7 nm respectively. All four molecules contain a thiol end group, which facilitates formation of high-quality SAMs on the (Ga,Mn)As [53,58]. The experimental details are described in the Methods section.

As described previously [53], despite the probable presence of direct contacts between the NM and (Ga,Mn)As (parallel conduction) through defects in the SAM, the spin valve conductance due to CISS in these junctions can be identified from the difference in junction conductance, ΔG, under opposite saturation magnetization for the (Ga,Mn)As. Figures 2a and 2b show representative sets of MC measurements with varying perpendicular magnetic field for (Ga,Mn)As/AHPA-L/NM junctions with NM electrode of Au and Al, respectively, measured at various constant bias currents at $T$ = 4.8 K. Each MC curve shows two distinct conducting states, coinciding with the well-defined square magnetic hysteresis of the (Ga,Mn)As due to its PMA, hence a ΔG can be precisely determined. The square hysteresis thus facilitates a straightforward and reliable determination of detailed bias current dependence of ΔG, from I-V curves measured under the opposite magnetization states of the (Ga,Mn)As, as shown in Fig. 2c. Specifically, the ΔG from the MC measurements can be obtained from and corroborated by the I-V's as [53]

$$\Delta G = I\left(\frac{1}{V_{-M}} - \frac{1}{V_{+M}}\right), \qquad (1)$$

where $V_{+M}$ and $V_{-M}$ indicate the corresponding bias voltage upon switching the magnetization in the (Ga,Mn)As from $+M$ to $-M$ at the same current $I$. Figure 2c shows ΔG as functions of bias current for the junction; as expected, the two types of measurements produced consistent results. The I-V data are presented in supplementary Fig. S1. The same measurement and analysis procedures were applied to all devices with different molecules and NM electrodes in this study to obtain ΔG and their bias current dependences.

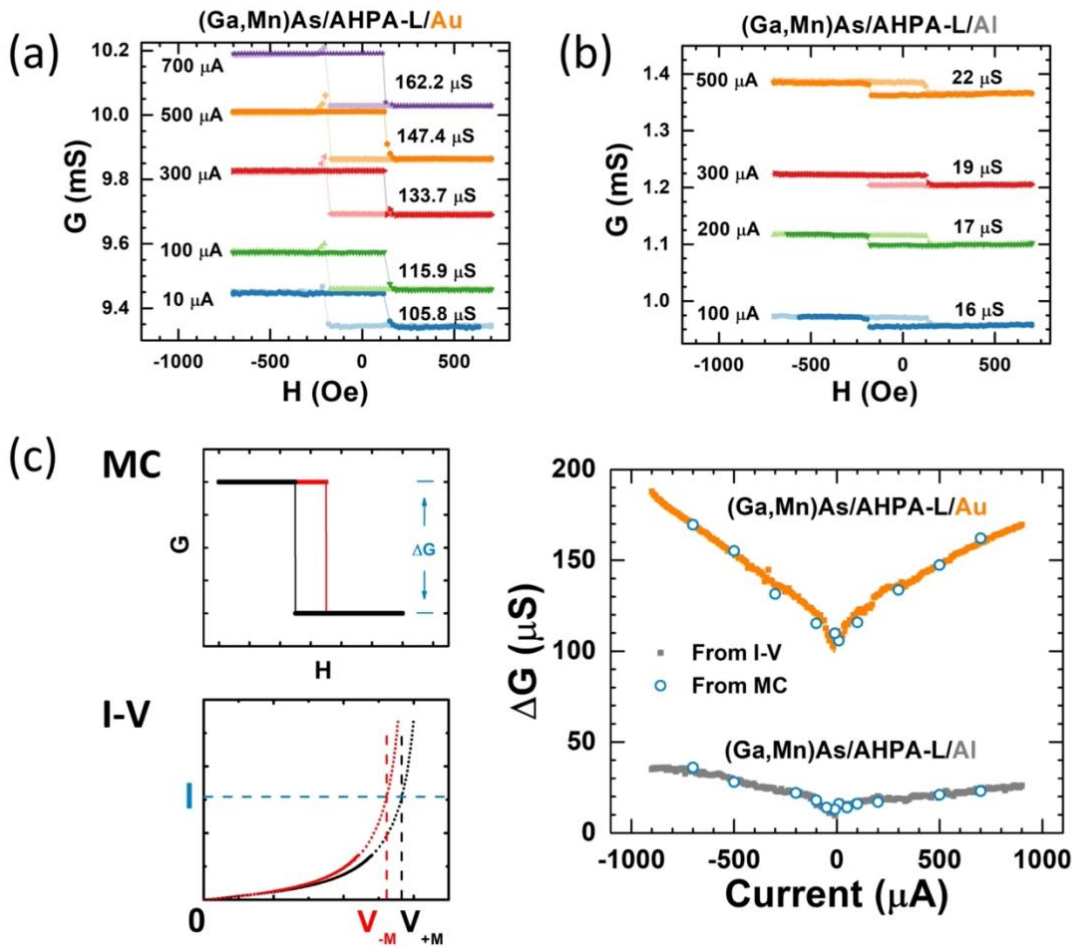

**Fig. 2: CISS spin valve conductance: effect of the NM electrode.** Representative MC curves measured at different bias currents for (Ga,Mn)As/AHPA-L/NM junctions, with NM of (a) Au, (b) Al. (c) The bias dependence of ΔG for the junctions with Au and Al contact. Pink squares are measured from I-V curves, whereas black circles are measured from MC measurements at different bias currents. The left panels illustrate how the value of ΔG are extracted from MC and I-V measurements.

We note that the total junction conductance (G) for the Au junction is also much greater than that of the Al junction. Nevertheless, measurements on the large number of (Ga,Mn)As/AHPA-L/Au devices clearly demonstrated that although the I-V and *total* conductance may vary greatly in the molecular junctions of similar structures depending on the degree of parallel conduction (quality of the SAM assembly), both the magnitude and bias current dependence of the CISS spin valve conductance (ΔG) were found to be consistently similar. A detailed discussion and comparison with a Au junction shown in our previous work [53] are presented in Supplementary Information (Sec. 2). We conclude that *the total G is spurious and has no bearing on CISS spin-valve conductance; it is the ΔG that accurately reflects the CISS effect*. The fact that different Au junctions show large variations of the I-V and total G, but exhibit similar bias-dependent ΔG, lends further credence to our model and the associated analyses and conclusion.

**Effect of the NM electrode**

Figures 2c shows CISS spin valve conductance for two (Ga,Mn)As/AHPA-L/NM junctions with Au and Al as the NM electrode. The experiment constitutes a direct comparison of the magnitude and bias dependence of ΔG for two NM electrodes of contrasting SOC strengths. The most notably result here is the pronounced differences between the junctions with Au and Al electrodes. Figure 3 shows the results from a comparative experiment with AHPA-L replaced by the much shorter chiral molecule of L-cysteine. As expected, with the same Au electrode, ΔG for the L-cysteine junctions are significantly smaller than those of the AHPA-L counterparts [52,59]. Remarkably, the large difference between ΔG is also present for the L-cysteine junctions with Au and Al electrodes. We emphasize that for each combination of chiral SAM and NM electrode, the entire set of measurements was repeated in multiple samples (2 to 4), each with 4 junctions, and the results were consistent. The experiments, therefore, strongly indicate that the observed significant impact of the NM electrodes on the CISS spin valve effect originates from inherent differences of Au and Al, and is independent of the specific chiral molecule used.

We have also fabricated and measured large numbers of devices with Cu and Ag electrodes, two NM materials of intermediate SOC strengths between Au and Al. However, despite the repeated attempts, for either material, we were unable to obtain results with the degree of consistency achieved in Au and Al devices. Most Ag and Cu junctions yielded very small ΔG without the regular bias dependences. We surmise that this was due to poor interfaces or even damages to the molecular SAM by Cu and Ag. As shown in supplementary Figure S3, one Cu junction yielded ΔG that fits well between those of the Au and Al junctions, however, it does not exhibit the bias-current dependence consistently seen in Au and Al junctions. For these reasons, we are unable to make a definitive statement regarding the CISS spin valve conductance and SOC strengths in Cu and Ag at this point.

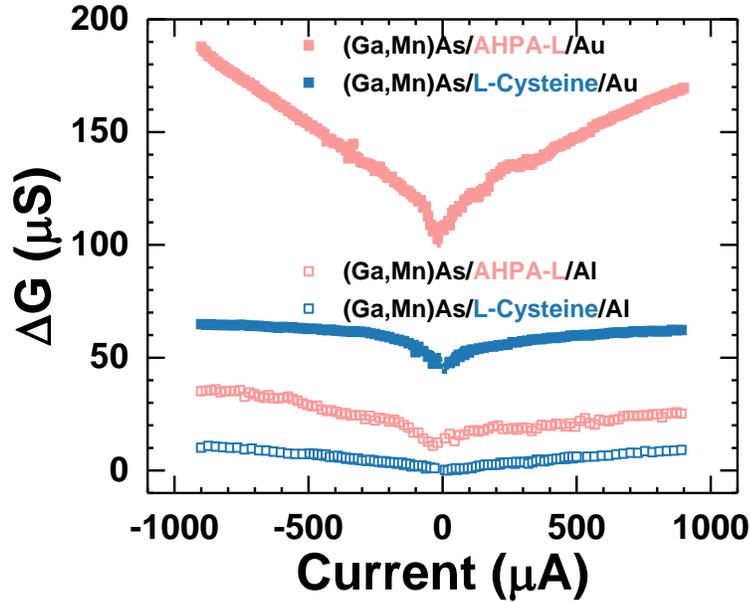

**Fig. 3: CISS spin valve conductance: chiral molecules of different lengths.** Bias current dependences of ΔG for four junctions of (Ga,Mn)As/AHPA-L (L-cysteine)/Au (Al).

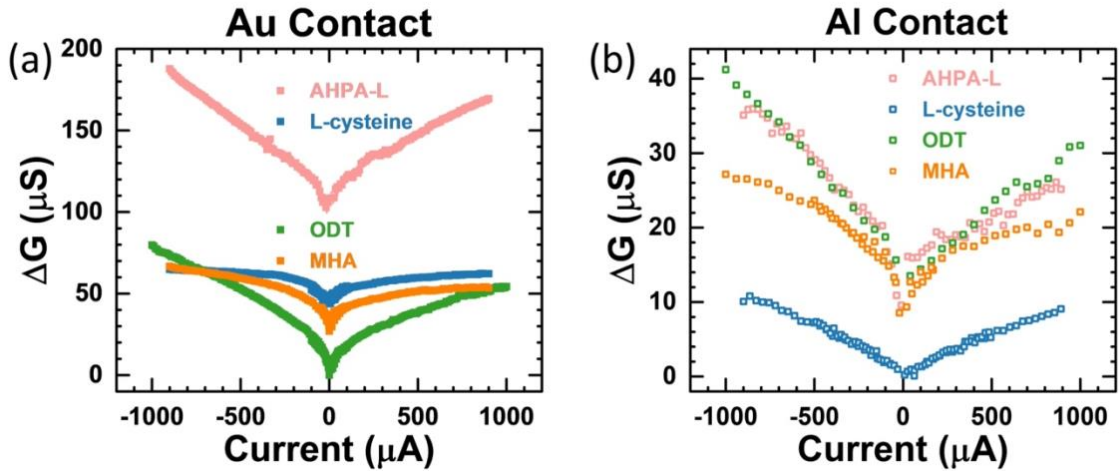

**Fig. 4: CISS spin valve conductance: chiral versus achiral molecules.** Bias current dependences of ΔG for the molecular junctions of different chiral (AHPA-L and L-cysteine) and achiral (MHA and ODT) molecules with (a) Au, and (b) Al contact. The pink and blue squares are the chiral molecular junctions (AHPA-L and L-cysteine), whereas green and orange squares are from the molecular junctions with achiral molecules (MHA and ODT).

**Chiral versus achiral molecules**

We further examined the spin valve effect in the molecular junctions with achiral molecules. Figure 4 shows the variation of the spin valve signal from chiral to achiral molecules in the molecular junctions with the same NM electrode of Au (Fig. 4a) and Al (Fig. 4b). In the Au junctions, there is a clear difference in the magnitude of the current-dependent ΔG between the chiral molecule (AHPA-L) and achiral molecules (MHA and ODT). It is important to note that despite the much-diminished magnitude, ΔG in the achiral molecular junctions are not trivial. Their magnitudes are clearly above the background in the control junction; in fact, they are comparable to ΔG in the short chiral molecule (L-cysteine) junctions. Moreover, ΔG in the MHA and ODT junctions exhibit distinct bias dependences resembling those in the chiral molecular junctions. These observations are consistent with a prediction from the orbital polarization model that nontrivial spin transport can materialize even in non-helical and even achiral systems, because in the presence of time-reversal symmetry, the emergence of orbital texture requires only inversion symmetry breaking [49], not necessarily chirality. Here, the Al junctions again provide an illuminating comparison (Fig. 4b): There are no discernible differences in the ΔG for the AHPA-L and achiral molecular junctions. In fact, as shown in Fig. S4, regardless of the molecule involved, the MC of all the Al junctions are comparable to that of the control junction without any molecules.

**Tunneling model and magnetochiral modulation of potential barrier**

Summarizing the key experimental observations, several robust features have been conclusively identified from our experiments: An optimal two-terminal chiral molecular spin valve, as exemplified by the (Ga,Mn)As/AHPA-L/Au junction, consistently exhibits significant MC, which increases linearly with the bias current at high biases but approaches a finite value at zero-bias. Contrary to expectation from the Onsager reciprocal relation [49,60,61], the MC is symmetric (i.e. the sign of Δ$G$ remains the same), rather than anti-symmetric, upon reversal of the current direction. Moreover, the magnitude of the MC decreases precipitously when AHPA-L is replaced by a much shorter chiral molecule or achiral molecules. Most importantly, replacing the Au electrode by Al results in even greater reduction of the MC, to levels where the differences between junctions of the different molecules are no longer discernible. Taken together, these results have revealed valuable new insights and placed several important constraints on a viable mechanism of CISS.

The orbital polarization model [49] offers a natural account for the observed qualitative differences between the molecular junctions with Au and Al electrodes. More recently, incorporating orbital polarization, a tunneling model was proposed to describe the electronic transport in chiral molecular junctions [56]. The essential ideas are depicted in Fig. 5a: The molecular chirality and electrode magnetism modulates tunneling barrier through the insulating molecular junction, termed magnetochiral modulation, which originates from the magnetochiral anisotropy [56]. We demonstrate that this model provides a semi-quantitative self-consistent description of all the key observations in the following.

We incorporate the magnetochiral modulation into the Simmons model [62,63] of metal/insulator/metal tunneling junctions. The problem of an arbitrarily shaped potential barrier is modeled into that of a rectangular barrier, which results in an explicit expression for the $I(V)$. The Simmons model and its variants have been widely applied to the modeling of electron transport in molecular junctions [64,65]. Based on the Simmons expression and assuming a small magnetochiral modulation of the potential barrier, in the intermediate bias range, we approximate the magnetization-dependent differential conductance through a chiral molecular junction in the form of a simple exponential relation:

$$\frac{dI}{dV} = (\alpha_o + \alpha_M)e^{\beta V_M} \quad (2)$$

where $\alpha_o$ and $\beta$ are magnetization-independent coefficients reflecting the tunneling current and probability across the unmodified potential barrier, and $\alpha_M$ is the coefficient that quantifies the effect from the change of the potential barrier height upon switching of the magnetization of the (Ga,Mn)As, namely the CISS spin valve conductance ΔG. The modulation of the potential barrier is small and considered a perturbation, $\alpha_M \ll \alpha_o$. ΔG as a function of the bias current (Eq. 1) can then be evaluated as:

$$\Delta G(I) = \beta I \left[ \frac{1}{\ln\left(1+\frac{\beta I}{\alpha_0+\alpha_{-M}}\right)} - \frac{1}{\ln\left(1+\frac{\beta I}{\alpha_0+\alpha_{+M}}\right)} \right] \quad (3)$$

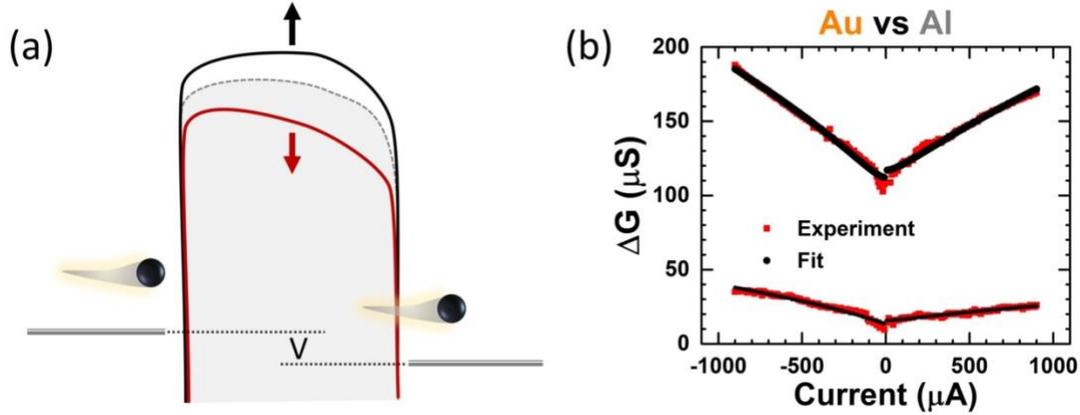

**Fig. 5**: **Barrier tunneling analysis of the CISS spin valve conductance.** The bias dependent ΔG for the chiral molecular junctions can be fitted to a model of magnetochiral modulation of the tunneling barrier. (a) Schematic depiction of the tunneling model and barrier modulation mechanism. Red and black curves illustrate the modified tunnelling barriers by ↓ and ↑ magnetizations, respectively, in the (Ga,Mn)As electrode. The original barrier is represented by the gray dashed curve. The bias voltage is indicated by V. (b) Fitting results of the spin valve conductance for two (Ga,Mn)As/AHPA-L/NM junctions with Au and Al contact.

Figure 5b shows the best fits of the ΔG for the two (Ga,Mn)As/AHPA-L/NM junctions (NM = Au, Al) to Eq. 3. More details of the fitting procedure are presented in Sec. 5 of the Supplementary Information. In brief, the fitting is performed separately for positive and negative currents. In a typical fitting process, an optimal value of $\beta$ is first identified. With $\beta$ fixed, best fit to the data to Eq. 3 is then performed with $\alpha_o$, $\alpha_{-M}$ and $\alpha_{+M}$ as the fitting parameters. Table 1 lists the resulting values for the parameters from the best fits.

| $\alpha$ value (µS) | Positive current | | | Negative current | | |
|---|---|---|---|---|---|---|
| Junction | $\alpha_o$ | $\alpha_{-M}$ | $\alpha_{+M}$ | $\alpha_o$ | $\alpha_{-M}$ | $\alpha_{+M}$ |
| (Ga,Mn)As/AHPA-L/Au | 1130 | 29.8 | -20.9 | 786 | 28.2 | -20.4 |
| (Ga,Mn)As/AHPA-L/Al | 836 | 2.92 | -3.95 | 233 | 2.42 | -3.49 |

**Table 1: Fitting parameters for the tunnelling model.** Values of $\alpha_o$, $\alpha_{-M}$ and $\alpha_{+M}$ from fittings of the CISS spin valve conductance data to Eq. 3. Here, β is kept constant at 10 V$^{-1}$. The most notable result is that for Au and Al junctions, $\alpha_o$ are similar while $\alpha_{\mp M}$ differ by an order of magnitude.

Two notable features are evident in Table 1. First, $\alpha_o \gg \alpha_{\mp M}$, consistent with our assumption that the magnetochiral modulation of tunneling barrier is small, and the CISS-induced spin valve conductance is a high-order effect in the electron transport. Furthermore, the magnitudes of $\alpha_{\mp M}$ in the junction with Au contact is an order of magnitude greater than those with Al electrode, while the values of $\alpha_o$ are similar in both junctions. The result constitutes quantitative support for the hypothesis that $\alpha_M$ is magnetization dependent and its strength depends on the SOC of the NM electrode. In addition, it is evident that the fittings provide excellent description of the observed bias current dependences of ΔG, and naturally account for the fact that ΔG is essentially independent of current direction. The higher order asymmetry in the data is reflected in the different values of $\alpha_o$ for positive and negative current. We note that slight asymmetries in the tunneling conductance are commonly observed and expected in the Simmons model [63] for junctions with dissimilar metallic electrodes and/or asymmetric potential barriers. The fitting results for other junctions, including those of achiral molecules, are described in Sec. 6 of the Supplementary Information. The results provide a quantitative measure of the different effects of the normal metal electrode (Au versus Al) consistent with the qualitative trends of ΔG apparent in Figures 2-4 and S4: For the Au junctions $\alpha_M$ are large and decrease significantly from AHPA-L to the achiral molecules, while all Al junctions show much smaller $\alpha_M$ without any systematic difference depending on the chirality and length of the molecules.

**Conclusions**

In summary, utilizing a robust device platform of magnetic semiconductor-based molecular junctions proven effective for CISS studies, we have obtained direct experimental evidence that the SOC in the NM electrode is essential to the emergence of the CISS spin valve effect. With a Au electrode, the precipitous decrease of the spin valve conductance from AHPA-L junctions to those of shorter chiral molecule and achiral molecules is readily discerned. Replacing the Au electrode with Al results in pronounced drops of the spin valve conductance for all molecules, to the degree that the differences between the molecules and with the control junctions are no longer discernible. A model based on orbital polarization from inversion-symmetry breaking and the resulting magnetochiral modulation of the tunneling barrier potential is shown to not only consistently account for all key aspects of the experimental results, but also provide resolution to several long-standing open issues in the field. Our work reveals an intimate relation between chirality and electronic properties, in which structural chirality information is transferred and transformed from molecular geometry to electronic orbital and eventually to the electronic spin via SOC. The results thus provide useful guidelines for detecting chirality-induced phenomena and designing CISS devices.

**Methods:**

1. **Materials**

The AHPA-L in the experiments was obtained from RS Synthesis, LLC. L-cysteine, MHA, and ODT were acquired from Sigma-Aldrich, Inc. Molecules, except L-cysteine, were dissolved in pure ethanol at 1 mM concentration. L-cysteine was dissolved in deionized water at a concentration of 2.5 mM. The AHPA-L solution was kept at -18 °C for storage whereas L-cysteine, MHA and ODT solution were stored in ambient conditions.

The (Ga,Mn)As with perpendicular magnetic anisotropy was grown by low-temperature molecular-beam epitaxy (LT-MBE) as described previously [53]. The (Ga,Mn)As thin film has a Curie temperature of 80 K and coercive field of 174 Oe.

2. **Device fabrication process**

The junction devices are fabricated in a similar process as previously reported [53]. The molecular assembly process is similar for both chiral and achiral molecules. During the deposition of the top magnetic electrode, the substrate is cooled with liquid nitrogen and temperature is maintained below -50 °C. 35 nm of Au (without Cr) or 50-70 nm of Al was deposited through a shadow mask.

3. **Electrical measurements**

All the measurements were performed in an Oxford $^3$He cryostat at the temperature range of 4.2 - 5.5 K. The measurement procedure is similar to previously reported [53] except that the perpendicular magnetic field is applied up to ±700 Oe with a constant rate of 350 Oe/min.

**Acknowledgement:**

We acknowledge helpful discussions with Hanwei Gao and Jiewen Xiao. The work at FSU is supported by NSF grant DMR-1905843. The work at IOS is supported by the MOST grant 2021YFA1202200 and the CAS Project for Young Scientists in Basic Research (YSBR-030). B.Y. acknowledges the financial support by the European Research Council (ERC Consolidator Grant "NonlinearTopo", No. 815869), the MINERVA Stiftung with the funds from the BMBF of Germany, and the Israel Science Foundation (ISF No. 2932/21).

# 1. I-V of molecular junctions with Au and Al electrode

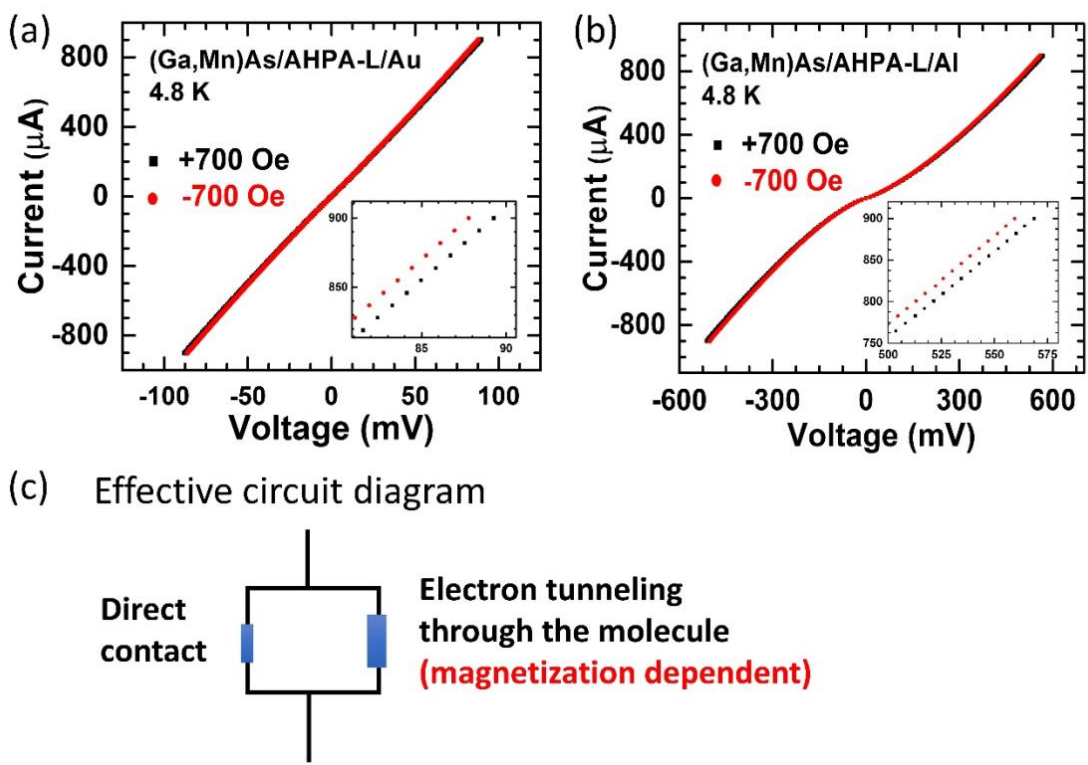

Figure S1: The I-V curves of the chiral molecular (AHPA-L) junctions with (a) Au and (b) Al contact in perpendicular magnetic fields of ±700 Oe. (b) The effective circuit diagram of a molecular junction showing parallel conduction of current through the direct contact of normal metal (NM) electrode and (Ga,Mn)As and through chiral molecules. Insets of (a) and (b) show the close-up images of the respective I-V's. (c) Effective circuit diagram with parallel conduction.

The field dependent I-V curves for two chiral molecular junctions with AHPAL SAM and Au and Al contacts are shown in Figure S1(a) and S1(b), respectively. The split between the I-V curves in perpendicular saturation fields of opposite polarities, ±700 Oe, is shown more clearly in the close-up images for the Au and Al junctions, respectively.

We note some apparent differences between the I-V curves for the Au junction here and a similar device in our previous work [1]: Here the I-V's are much more linear and the bifurcation is much less obvious. We believe the differences stem from a larger contribution to the current from the parallel contribution through direct contact between the Au and (Ga,Mn)As, as illustrated in Fig. S1(c).

## 2. Spin valve conductance (ΔG) versus total junction conductance (G)

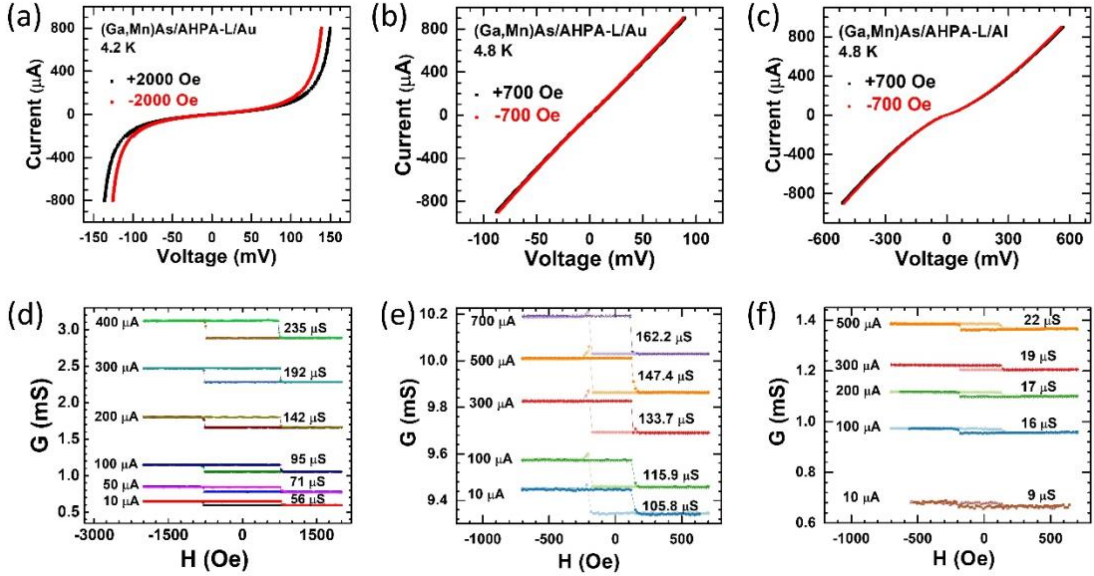

Figure S2: (a), (b), (c) I-V curves for a (Ga,Mn)As/AHPA-L/Au junction from our previous work [1], and a (Ga,Mn)As/AHPA-L/Au junction and a (Ga,Mn)As/AHPA-L/Al junction in the present work, respectively. (d) - (e) show the corresponding MC measurements for the three junctions.

As was elaborated previously [1] and emphasized again in the main text here, because of the contribution to the total current from the parallel conduction [Fig. S1(c)], the I-V and **total** conductance of the molecular junctions of similar structures may vary greatly depending on the degree of parallel conduction (quality of the SAM assembly), however, both the magnitude and bias current dependence of the CISS spin valve conductance (ΔG) of the junctions are found to be consistently similar. This is well illustrated in Fig. S2, which shows the I-V and magnetoconductance measurements for a (Ga,Mn)As/AHPA-L/Au junction in our previous work [1], and the (Ga,Mn)As/AHPA-L/Au(Al) junctions in this work. *Clearly, the zero-bias total junction conductance of the Au junction in the previous work is much **smaller** than that of the Au junction in this work, and even smaller than that of the Al junction here, whereas the ΔG for that junction are **comparable** to those of the Au junction in the present work at low bias currents (e.g., 100 μA), and becomes significantly **greater** at high biases.* We believe the more nonlinear I-V's and smaller total junction conductance in the Au junction in the previous work are signatures of a smaller contribution to the total current from the parallel contribution through direct contact between the Au and (Ga,Mn)As. This conclusion is also corroborated by the more rapid increase of ΔG with the bias current seen in the first Au junction.

In summary, these observations are compelling evidence that *the total G is spurious and has no bearing on CISS spin-valve conductance; it is the ΔG that truly reflects the CISS effect*. The fact that different Au junctions show large variations of the I-V and total G, but exhibit similar bias-dependent ΔG, lends further credence to our model and the associated analyses and conclusion.

## 3. Other normal metal (NM) electrode materials of varying SOC strengths

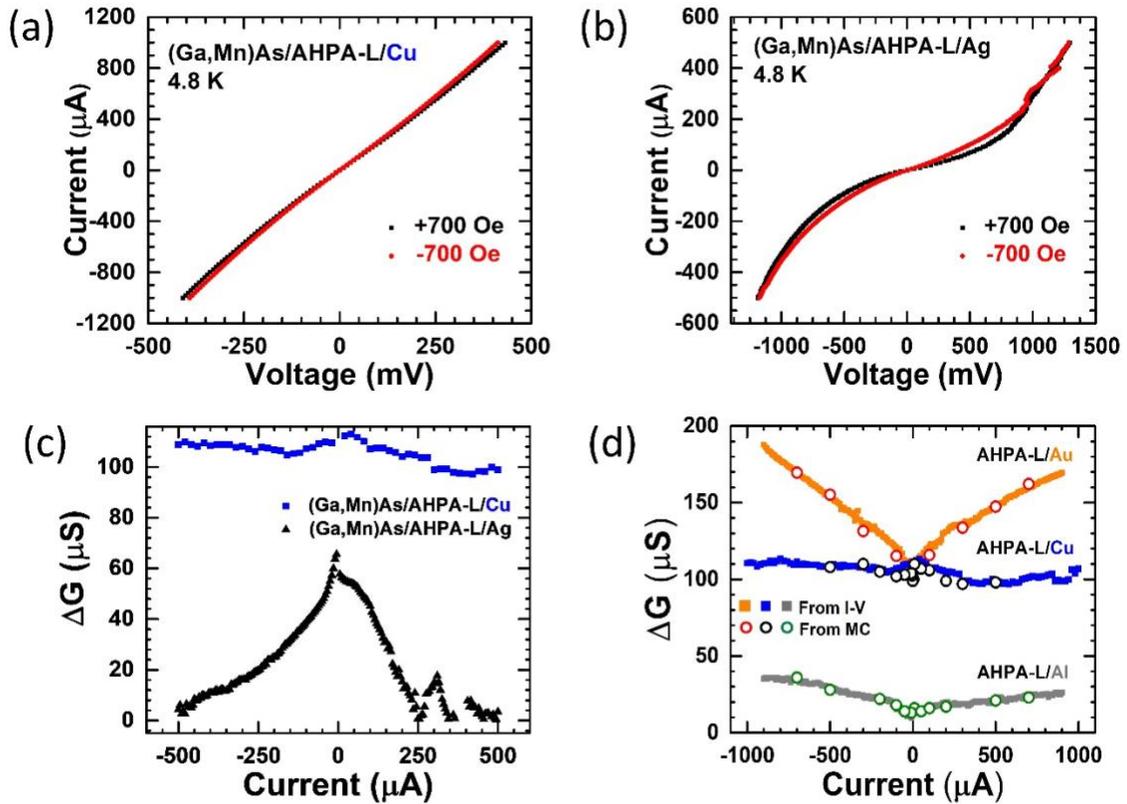

Figure S3: The I-V curves for chiral molecular (AHPA-L) junctions with (a) Cu, and (b) Ag, contact in perpendicular magnetic fields of ±700 Oe. (c) The bias dependence of ΔG for the junctions with Cu and Ag contact extracted from the I-V curves. (d) Comparison of the bias-current dependent ΔG for the junctions with Au, Cu. and Al contact. The solid squares are ΔG extracted from I-V curves, and the open circles are ΔG from MC measurements at different bias currents.

We have fabricated and measured large numbers (>10) of devices with Cu and Ag electrodes, two NM materials of intermediate SOC strengths between Au and Al. However, despite the repeated attempts, for either material, we were unable to obtain results with the degree of consistency seen in Au and Al devices. We speculate that this originated from poor interfaces of the molecular SAM with Cu and Ag, possibly due to the higher evaporation temperatures or some specific chemistry with the organic molecules for Cu and Ag. Figure S3 shows a representative I-V and resulting ΔG from a Ag junction, and a **best** set from the Cu junctions. Although the magnitude of ΔG for this Cu junction falls well between those of the Au and Al junctions, as shown in Fig. S3(d), it does not exhibit the bias-current dependence consistently seen in Au and Al junctions. Combined with the large sample to sample variations, we are unable to make a definitive statement regarding the CISS spin valve conductance and SOC in Cu, in contrast to Au and Al.

## 4. Comparison with control samples without molecules

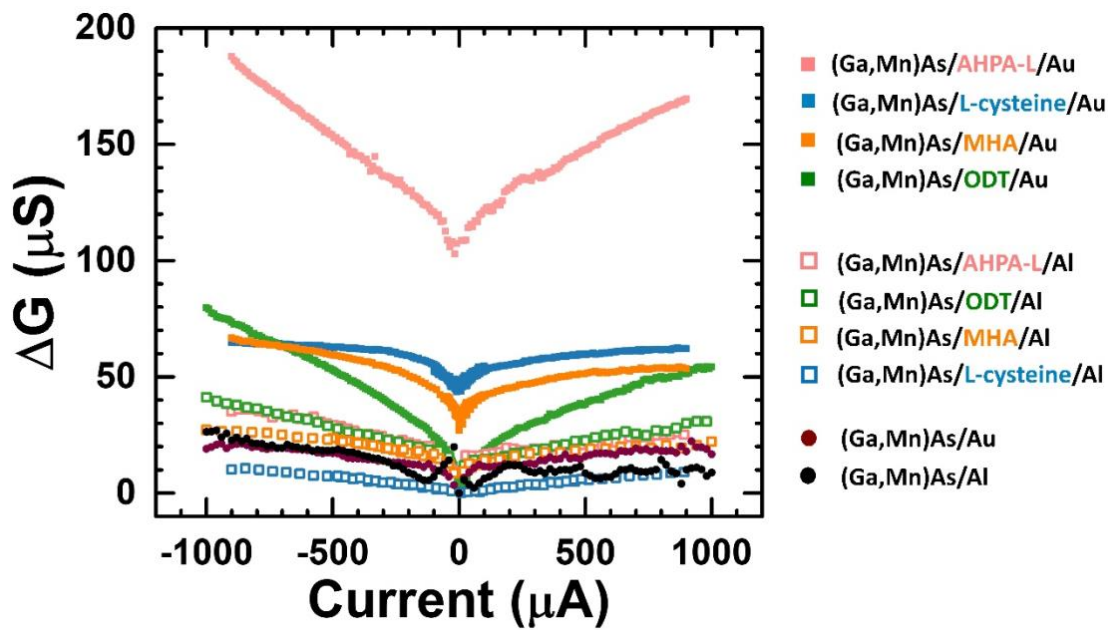

Figure S4: Comparison of the bias-dependent spin valve conductance (ΔG) of molecular junctions (chiral and achiral) with Au and Al normal metal electrodes, and that of control junctions without any molecules with Au and Al electrodes.

The control samples were fabricated with the same device fabrication process omitting the assembly of molecules. Figure S4 shows the comparison of the bias dependences of the spin valve MC of molecular junctions (chiral and achiral) with Au and Al normal metal contacts, and those of control samples with Au and Al contacts on (Ga,Mn)As without any molecules. It is worth noting that the bias-dependent MC of the molecular junction with Al electrode is similar to that in the control junctions with both Au and Al contact, suggesting that the MC of molecular junctions with Al electrode is essentially negligible.

## 5. Further details on the fitting procedure

The fitting of the bias dependence of $\Delta G$ using Eq. (3) in the main text was done both in Origin and by using the curve fitting toolbox (cftool) of MATLAB. The fitting was performed separately for positive and negative bias currents, using $\alpha_o$, $\alpha_{-M}$ and $\alpha_M$ as fitting parameters. The value of $\beta$ was kept constant at 10 V$^{-1}$ while fitting bias dependence of both chiral and achiral molecules. The typical process of fitting in Origin along with the parameter values, standard errors, and Adj. R-Square are shown in Fig. S5. We point out that the resulting values of the parameters are similar for fittings done by Origin and MATLAB.

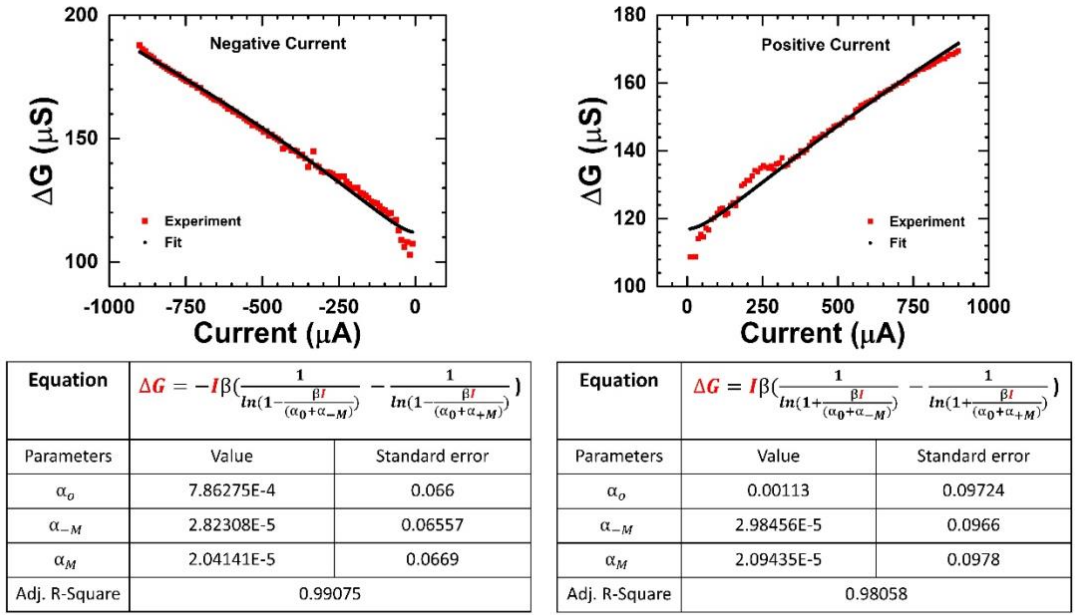

Figure S5: An example depicting the process of curve fitting of the bias-dependent ΔG in Origin. The data is from the spin-valve conductance for a (Ga,Mn)As/AHPA-L/Au junction. Similar process of fitting was employed for all molecular junctions with Au and Al contact.

## 6. Fitting results for junctions of achiral molecules

Using the procedure described in the previous section, fittings to Eq. (3) were performed for the achiral molecular junctions (MHA and ODT) with Au and Al normal metal electrodes. Figure S6 shows the best fits. Table S1 lists the resulting values for the parameters from the best fits.

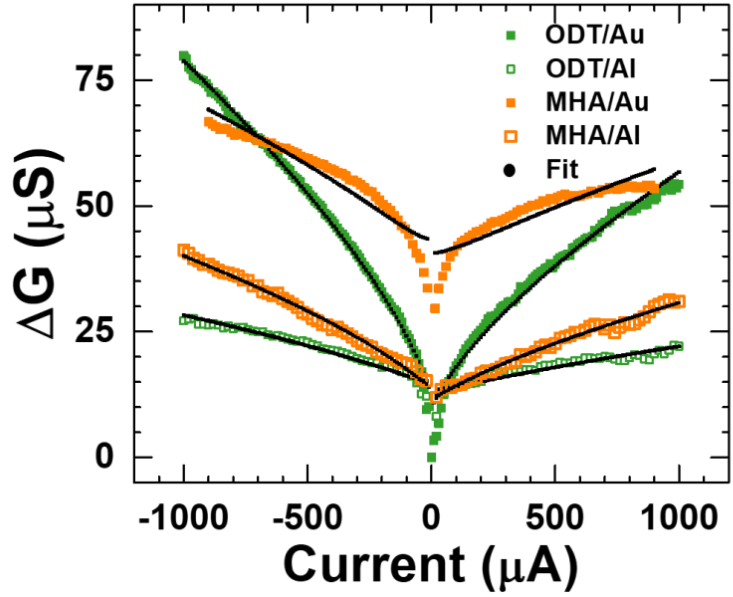

Figure S6: Fitting results of the spin valve conductance for (Ga,Mn)As/achiral (MHA and ODT)/NM junctions with Au and Al contact.

| Achiral molecules (MHA and ODT) | | | | | | |
|---|---|---|---|---|---|---|
| Junction | Negative current | | | Positive current | | |
| | $\alpha_o$ | $\alpha_{-M}$ | $\alpha_M$ | $\alpha_o$ | $\alpha_{-M}$ | $\alpha_M$ |
| ODT/Au | 98.6 µS | 8.63 µS | -6.72 µS | 120 µS | 7.52 µS | -6.83 µS |
| ODT/Al | 312 µS | 3.80 µS | -2.74 µS | 421 µS | 2.31 µS | -3.53 µS |
| MHA/Au | 676 µS | 8.84 µS | -9.97 µS | 831 µS | 8.16 µS | -9.47 µS |
| MHA/Al | 254 µS | 1.89 µS | -3.29 µS | 305 µS | 2.13 µS | -3.07 µS |

Table S1: Fitting parameters of the CISS spin valve conductance for the achiral molecular junctions with Au and Al contacts.

The values of the parameters for spin valve conductance of achiral molecular junctions obtained from the fitting are consistent with our expectation. The notable aspect of this fitting is that the values of $\alpha_{\mp M}$ in the junction with Au are much greater than those with Al, even though the values are in the same order. This is consistent with the observation of diminished but finite spin valve conductance in achiral molecules with Au contact. The values of $\alpha_o$ are similar to that of chiral molecular junctions.

**References:**

[1] T. Liu, X. Wang, H. Wang, G. Shi, F. Gao, H. Feng, H. Deng, L. Hu, E. Lochner, P. Schlottmann, S. von Molnár, J. Zhao, and P. Xiong, "Linear and Nonlinear Two-Terminal Spin-Valve Effect from Chirality-Induced Spin Selectivity", *ACS Nano* 14, 15983 (2020).